\documentclass[conference]{IEEEtran}
\IEEEoverridecommandlockouts
\usepackage{cite}
\usepackage{amsmath,amssymb,amsfonts}
\usepackage{algorithmic}
\usepackage{graphicx}
\usepackage{subcaption}
\usepackage{textcomp}
\usepackage{xcolor}
\usepackage{multirow}
\usepackage{algorithm}
\usepackage{authblk}
\usepackage{amsmath, amssymb}
\usepackage{tabularx}
\usepackage{array}
\usepackage{stfloats}
\def\BibTeX{{\rm B\kern-.05em{\sc i\kern-.025em b}\kern-.08em
    T\kern-.1667em\lower.7ex\hbox{E}\kern-.125emX}}
\begin{document}

\title{From Agent Simulation to Social Simulator: A Comprehensive Review (Part 2)\\
}
\renewcommand\Affilfont{\small}
\author[1,2,3*]{Xiao Xue}
\author[4,5]{Deyu Zhou}
\author[6]{Ming Zhang}
\author[1,2,3]{Xiangning Yu}
\author[7]{Fei-Yue Wang}

\affil[1]{\textit{College of Intelligence and Computing, Tianjin University, Tianjin, China}}
\affil[2]{\textit{Tianjin Key Laboratory of Healthy Habitat and Smart Technology, Tianjin, China}}
\affil[3]{\textit{Laboratory of Computation and Analytics of Complex Management Systems, Tianjin University, Tianjin, China}}
\affil[4]{\textit{School of Software, Shandong University, Jinan, China}}
\affil[5]{\textit{Joint SDU-NTU Centre for Artificial Intelligence Research (C-FAIR), Shandong University, Jinan, China}}
\affil[6]{\textit{Faculty of Environment, Science and Economy, University of Exeter, Exeter, UK}}
\affil[7]{\textit{Institute of Automation Chinese Academy of Sciences, Beijing, China}}

\affil[ ]{ }
\affil[ ]{Email: jzxuexiao@tju.edu.cn, zhoudeyu@mail.sdu.edu.cn, mz427@exeter.ac.uk, yxn9191@gmail.com, feiyue@ieee.org}

\maketitle

\begingroup
\renewcommand\thefootnote{}\footnotetext{*Corresponding author: Xiao Xue, e-mail:jzxuexiao@tju.edu.cn.\\

**This paper is part of a planned multi-part review series on From Agent Simulation to Social Simulator.

The present article constitutes Part II of the series. It is self-contained and can be read independently; readers are not required to have read Part I in order to understand the material presented here.

While the authors have previously published related work in IEEE journals, \textit{IEEE Transactions on Computational Social Systems} (DOI: 10.1109/ICWS67624.2025.00068) and \textit{IEEE Journal of Automatica Sinica} (DOI: 10.1109/JAS.2024.124221), those earlier studies were conducted under a different technological context. Motivated by recent advances, this paper offers a new organization, scope, and synthesis of the topic. This article is not a revision or updated version of any previously published work.

}
\endgroup

\begin{abstract}
The study of system complexity primarily has two objectives: to explore underlying patterns and to develop theoretical explanations. Pattern exploration seeks to clarify the mechanisms behind the emergence of system complexity, while theoretical explanations aim to identify the fundamental causes of this complexity. Laws are generally defined as mappings between variables, whereas theories offer causal explanations of system behavior \cite{wang2010emergence, yilma2021systemic, salmon1998causality}. Agent Based Modeling (ABM) is an important approach for studying complex systems \cite{gilbert2005simulation}, but it tends to emphasize simulation over experimentation. As a result, ABM often struggles to deeply uncover the governing operational principles. Unlike conventional scenario analysis that relies on human reasoning \cite{morgan2014counterfactuals}, computational experiments emphasize counterfactual experiments—that is, creating parallel worlds that simulate alternative “evolutionary paths” of real-world events \cite{xiao2023putational}. By systematically adjusting input variables and observing the resulting changes in output variables, computational experiments provide a robust tool for causal inference, thereby addressing the limitations of traditional ABM. Together, these methods offer causal insights into the dynamic evolution of systems. This part can help readers gain a preliminary understanding of the entire computational experiment method, laying the foundation for the subsequent study.
\end{abstract}


\section{The Laws of System Complexity}
Complexity is one of the most intricate concepts in modern science \cite{mitchell2009complexity, simon2012architecture}, and no unified definition has yet been established; perhaps such a definition may never exist. The different layers of the real world—physical, biological, social, and conscious layers—each exhibits distinct forms of complexity that should not be conflated. One cannot substitute lower-level complexity for higher-level complexity, nor dismiss lower-level complexity on the basis of higher-level complexity. Complexity can be understood from various perspectives and at different levels. This section first provides a categorical overview of the causes of complexity, followed by a detailed discussion using structural complexity and distributional complexity as examples. The formulas presented here are mainly drawn from Claudio Cioffi-Revilla’s work Introduction to Computational Social Science: Principles and Applications. For more details, please refer to the original text \cite{cioffi2014introduction}.

\subsection{Emergent Phenomena and Causal Identification Challenges in Complex Systems}
To explore the causal mechanisms of complex systems \cite{holland2000emergence}, it is first necessary to understand their emergent phenomena. Jochen Fromm’s classification \cite{fromm2004emergence} offers profound insights into these emergent behaviors. As illustrated in \textbf{Figure 1}, emergent phenomena can be categorized into four types based on their causal relationships.

The first type is simple emergence, which involves unidirectional, bottom-up causality and is typically reflected in basic statistical properties. In contrast, weak emergence is more intriguing because it reflects top-down causality, where individual behaviors are influenced by collective rules, leading to expressions of collective intelligence. Multiple emergence captures the intricacies of interactions between different causal layers, as seen in simulations like the “Game of Life,” where static and dynamic patterns of various sizes and shapes coexist, demonstrating the interplay of multi-level causal relationships. The most complex form, strong emergence, encompasses three levels: micro, meso, and macro.

Understanding these types of emergence is crucial for causal inference in complex systems. While macro-level phenomena originate from micro-level mechanisms, they can effectively substitute for micro-level descriptions in causal reasoning. This is the core idea of causal emergence. Macro-level behavioral descriptions are often more intuitive, whereas micro-level mechanisms contain the full complexity of the system. However, how micro-level causal relationships relate to those at the macro-level remains an open question, requiring further exploration and also involving addressing the following challenges:

\begin{figure}[htbp]
\centering
\includegraphics[width=\linewidth]{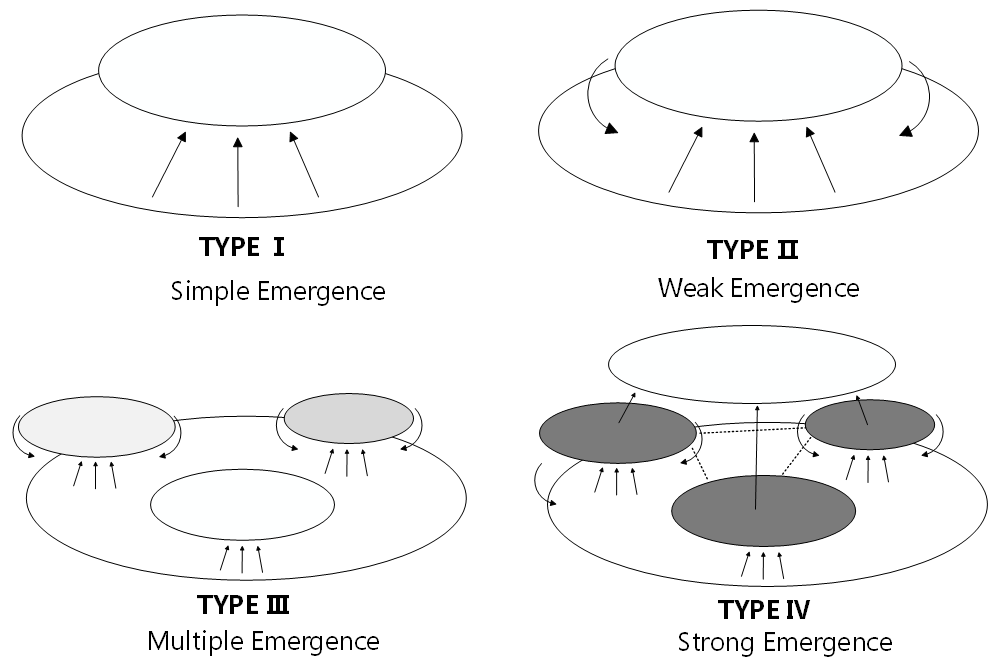}
\caption{Fromm’s Classification of Emergence.}
\end{figure}

\textbf{Challenge 1: Exploring the Interrelationship between Multiple Influencing Factors and Complexity.} 

Complex systems involve a series of complex factors and their interwoven relationships, including disturbances from both external environmental factors and internal fluctuations or human-induced errors. Such noise obscures experimental data, making accurate causal inference difficult. Noise not only hides true causal relationships but also risks leading to incorrect conclusions.

\textbf{Complexity Arising from System Scale:} The number of system components serves as an indicator of the system’s scale. The formation of complexity requires a sufficiently large system size. A vast scale poses difficulties for description and processing, rendering traditional methods ineffective. Simple systems do not exhibit complexity. A sufficiently large scale is a necessary condition for the generation of complexity, but it is not a sufficient condition. Even a giant system is not necessarily a complex system. 

\textbf{Complexity Arising from System Structure:} The diversity and heterogeneity of system components lead to a variety of interrelationships and disparities, which are fundamental drivers for the emergence of complexity. In terms of complexity formation, structural effects are more significant than scale effects. The greater the heterogeneity of components, the harder it is to integrate them. Notably, hierarchical structures are one of the main sources of complexity, and researchers in the field of complexity almost unanimously emphasize this point. Non-hierarchical structures, which are relatively simple, have only two levels: elements and the whole. Hierarchical structures involve intermediate layers between elements and the system as a whole. Without navigating and integrating different levels, the final system cannot be fully constructed. The more layers there are, the more likely complexity will emerge.

\textbf{Complexity Arising from Openness (Environment):} Closed systems do not exhibit complexity; complexity must arise in open systems. When external environmental factors no longer act as disturbances and the system’s inherent structure instead becomes dominant, closed systems face two issues:  they cannot resist external shocks, and black-box methods fail to capture their dynamics. Consequently, such systems inevitably show certain types of complexity. Openness is a major contributor for the emergence of complexity, and the interaction between complex systems and their environments is a critical manifestation of systemic complexity\cite{prigogine2018order}.

\textbf{Challenge 2: Agent Learning and Adaptability Increases the Complexity of Causal Identification.}

\textbf{Complexity Arising from Proactivity and Dynamism:} In simple systems, actors are passive, and causes and effects are clearly distinguishable. In contrast, in complex systems, different components interact and co-evolve with their environments, forming networks of causal interdependencies that can lead to complexity, especially when agents possess a certain level of autonomy and continuously adapt to their environment. In this process, the system as a whole may exhibit emergent behavioral complexity. A core tenet of research at Santa Fe Institute research is that adaptivity gives rise to complexity. All adaptive systems (i.e., complex adaptive systems, CAS) emerge through continuous adaptation to dynamic environments.

\textbf{Complexity Arising from Individual Intelligence:} Even if a system composed of non-intelligent individuals generates complexity through self-organization, it is usually relatively rudimentary and low-level. Intelligent agent systems (such as those studied by the Santa Fe Institute’s complex adaptive systems research) can sense their surrounding environments, predict future system states, learn from experience, and develop behavioral rules, thereby achieving adaptive transformations. Thus, they are inherently complex. The higher the intelligence of individual agents, the greater the system’s complexity. Agent intelligence is one of the major sources of complexity. For example, land combat systems typically involve a limited number of agents but are still complex due to the high intelligence of each agent, resulting in unpredictable and emergent system behaviors. These systems are therefore classified as open complex systems.

\textbf{Complexity Arising from Human Rationality:} In systems where humans are essential components, system behavior is influenced by human rationality. In particular, in competitive systems such as games, the rationality of participants (e.g., strategic thinking, decision-making logic, risk assessment) is a significant source of complexity. If we assume complete rationality (i.e., unlimited reasoning), the root of complexity may be eliminated, and game players would always adopt the “max-min strategy.” Such systems are simplified and can be analyzed using standard game theory. However, bounded rationality (i.e., limited reasoning) leads to the emergence of true complexity\cite{simon1955behavioral}.

\textbf{Challenge 3: Emergence from Micro to Macro Levels Masks the Underlying Micro Mechanisms.}

\textbf{Complexity Arising from Nonlinearity:} Philosophers have long pointed out that the ultimate cause for the development and change of things is interactions, but interactions can be either linear or nonlinear. Linearity implies additivity, uniformity, invariance. It fails to meet the conditions for generating complexity. Linear systems are usually simple; linear interactions produce predictable outcomes and cannot give rise to complexity. Nonlinearity, by contrast, implies rich diversity, heterogeneity, variability, asymmetry, irregularity, singularity, and creativity. Nonlinear interactions between elements or subsystems are the internal mechanisms behind complexity. However, linear systems themselves can also contain significant heterogeneity. Weak nonlinearity may not necessarily lead to complexity, and it is often treated as a disturbance term. In such cases, linear models supplemented with perturbation methods may still describe the system effectively. Only strong nonlinearity, especially intrinsic nonlinearity, can truly give rise to complexity\cite{kauffman1992origins}.

\textbf{Complexity Arising from Uncertainty:} Randomness is a key form of uncertainty. Simple systems exhibit stable random processes, which can be modeled statistically. Physical systems with well-behaved random processes often conform to the law of large numbers and fall within the scope of statistical analysis. However, in living systems and social systems—where components are intelligent and interact in complex ways—relying solely on statistical laws cannot capture the system’s fundamental characteristics. Current statistical approaches are inadequate to handle such randomness. For example, it is impossible to deduce the overall behavior of a biological system solely from individual genetic markers, nor can the overall characteristics of a social system be statistically inferred from features of individual members. Another major source of uncertainty is fuzziness. Fuzziness is both a cause of complexity and a result of its expression. Fuzzy set theory was introduced specifically to deal with such complexity. It holds that when system complexity surpasses a certain threshold, precision in description and the value of interpretation are often incompatible; the two cannot coexist.

\subsection{System Complexity Theory: Explanation}
A fundamental function of theory is to explain phenomena. Therefore, a theory must account for observed events by drawing on initial conditions or driving factors, and its arguments must conform to the model of scientific explanation. A defining feature of scientific theories is that they must present a narrative linking antecedents (causes) with consequences (effects)—a causal story. System complexity theory explains the emergence of complexity in systems through causal relationships. Its distinguishing feature lies in its ability to interpret observed facts or empirical patterns using the available data.

\begin{table*}[htbp]
\centering
\caption{Definitions of Related Concepts in System Complexity.}
\renewcommand{\arraystretch}{1.4}
\begin{tabular}{|>{\raggedright\arraybackslash}p{4cm}
                |>{\raggedright\arraybackslash}p{6.5cm}
                |>{\raggedright\arraybackslash}p{6cm}|}
\hline
\textbf{Concept Name} & \textbf{Concept Explanation} & \textbf{Remarks} \\
\hline

System Complexity Emergence &
Suppose system complexity is represented as a composite event $C$ at the macro level. It originates from the combination of a set of basic events (sample points in the sample space $\Omega$) and the coupling process among these basic events. &
The composite event $C$ consists of two components: (1) a series of basic observational events (defined by coupling outcomes and relevant natural state structures); (2) a rule of operations that links these events together. \\
\hline

Event Function &
When the emergence of a complex event $Y$ is related to the occurrence or sequence of other events $\mathbf{X}$, the mapping $\varphi: (\mathbf{X}) \rightarrow Y$ is referred to as the event function of $Y$. &
The event function $\varphi()$ defines causal explanation and essentially corresponds to a functional mapping for causal inference. Thus, the event $Y$ is expressed as $Y = \varphi(\mathbf{X})$. \\
\hline

Emergence Event Function of System Complexity &
When the emergence of system complexity, represented by a composite event $C$, is related to the occurrence or sequence of other events $\mathbf{X}$, the mapping $\varphi: (\mathbf{X}) \rightarrow C$ is referred to as the emergence event function of $C$. &
Therefore, the event $C$ is expressed as $C = \varphi\{\mathbf{X}\}$. \\
\hline

\end{tabular}
\end{table*}

Given that system complexity emerges as a result of individual-level games (as opposed to most natural state outcomes), the modeling and explanation of the emergence of system complexity depend on the outcomes of these individual strategic interactions. Based on whether causal events occur or not within the sample space, the emergence of system complexity can be interpreted through the causal logic of event occurrence. That is, causal events must occur in a non-random manner in order for system complexity to manifest. For a collective behavior to emerge, certain key combinations of causal events must occur in a specific configuration; otherwise, the collective behavior will not emerge. For instance, the formation of an international alliance requires that events related to national strategy, leadership capacity, interest alignment, and coalition recruitment occur in a particular configuration, or in an equivalently effective alternative configuration.

Formally, the event function can precisely represent the causal logic and explain how a composite event arises. However, this raises several questions: What types of event functions exist? How do different event functions explain the emergence of composite events? How can the probability of a composite event be determined from an event function? To address these and similar questions, we must examine the logic of system complexity at the micro level and identify two causal explanatory models based on sequential logic and conditional logic. According to the conceptual framework of Simon’s theory \cite{newell1972human}—including environmental complexity, goal-seeking behavior, bounded rationality, adaptiveness, and near-decomposability—the sequential tree of forward causal logic enables the construction of  models of adaptation and system complexity. The model shown in \textbf{Figure 2} provides a first-order representation of Simon’s theory. The key insight is that various outcomes within the sample space $\Omega$ can be generated through combinations. Specifically, the emergence of system complexity C requires at least four necessary sequential conditions, which collectively have a notably low probability; without them, the successful emergence of system complexity C is impossible. In contrast, alternative outcomes (failures $E$, $E^{*}$, and $E^{**}$) occur with relatively higher probabilities, making them more likely to manifest overall.

\begin{figure}[htbp]
\centering
\includegraphics[width=\linewidth]{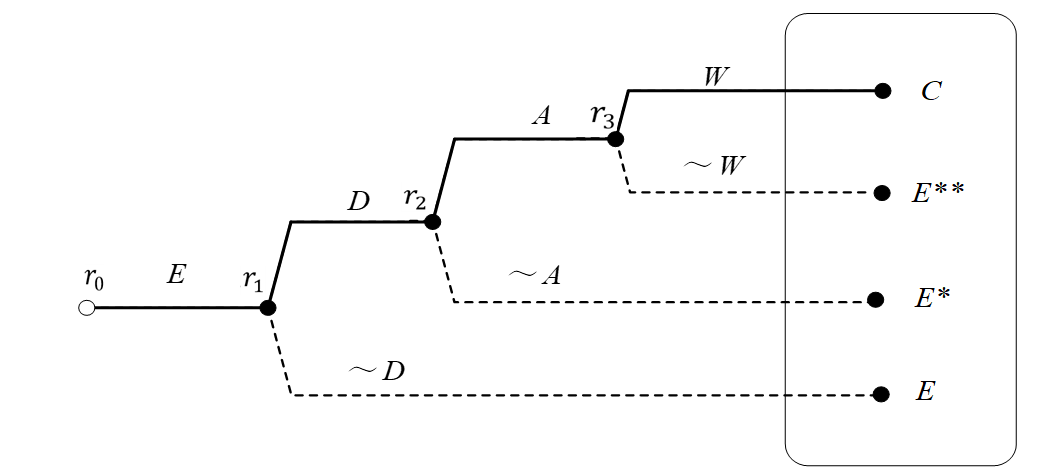}
\caption{Forward Sequential Causal Logic Tree of Simon’s Adaptation Theory and the Emergence of System Complexity.}
\end{figure}

At certain initial time points $r_0$, event E occurs within a specific environment. At subsequent time points $r_1$, in response to environmental challenges, individuals can decide whether to make adaptive changes based on bounded rationality, i.e., event D. If they choose not to change (i.e., event ~D), they will continue to endure the same environmental impacts as before, regardless of the outcome (result E).

If they decide to adapt, they then face the choice of whether to actually implement the decision and carry out adaptive measures—this is event A. If they fail to execute this action (i.e., event ~A), they must still endure the environmental impact; however, time has passed, leading to outcome $E^{*}$. It can be argued that $E \approx E ^{*}$. 

If individuals have taken action, at some point in time, behavioral feedback may or may not be effective. If the feedback is effective (event W), the result is success with increased complexity, since a more complex system must now be maintained (result C). If the feedback is ineffective (event ~W), the result is failure with continued exposure to environmental impact (result $E ^{**}$). It can be argued that $S(E^{**}) > S(E)$, where $S(X)$ denotes the impact or inefficacy associated with event $X$.

\section{Causal Inference Framework of Computational Experiments}
The nonlinear emergent characteristics of complex systems present significant challenges to causal inference. A core issue in computational experimental approaches is understanding in what sense and under what conditions agent-based models (ABMs) can effectively support causal reasoning. ABMs are capable of simulating sequences of events generated by low-level or small-scale interactions, and of gradually generating the outcomes associated with those sequences. In this way, they provide a means to explicitly model the causal mechanisms that may give rise to observed dependencies over time. However, despite their potential, ABMs often fall short in terms of generating empirical evidence or compelling arguments that convince audiences that the modeled mechanisms truly reflect those found in the real world. This section therefore reviews the intersection of ABMs and causal inference, aiming to clarify the nature of the relationship between ABMs and causal inference.

\subsection{Horizontal Causation and Vertical Causation}
In the study of causality within complex systems, Hoel et al.’s framework of causal emergence \cite{hoel2013quantifying} highlights that micro-level dynamics are often noisy, leading to weaker causal relationships at that scale, whereas macro-level causal relationships are generally stronger and more discernible. The micro-level refers to individual agents interacting according to specific rules, while the macro-level encompasses the collective behaviors or emergent properties arising from those interactions. These emergent macro-level properties cannot be directly inferred from any single micro-level component, as they are the result of intricate interaction processes. This suggests that a system’s overall behavior can be more complex and less predictable than the mere sum of its parts. While many current studies \cite{klein2020emergence, hoel2013quantifying, shalizi2003macrostate, barnett2023dynamical} focus on macro-level analysis, examining how inputs affect outputs, they often fail to illuminate the underlying causal mechanisms, which are crucial to understanding emergent phenomena.

Casini and Manzo, drawing on Hall’s classification framework, divide causal theories into two categories: horizontal causation (causal inference) and vertical causation (causal explanation) \cite{casini2016agent}. From the horizontal perspective, causal mechanisms are understood as networks of variables in which a cause variable influences an outcome variable through a chain of intervening variables. In contrast, the vertical perspective conceptualizes causal mechanisms as “complex system” \cite{glennan2002rethinking}, where the behavior (or property change) of one individual can trigger behaviors or property changes in others, and interactions among micro-level individuals give rise to system-level phenomena or behaviors.

\begin{figure}[htbp]
\centering
\includegraphics[width=\linewidth]{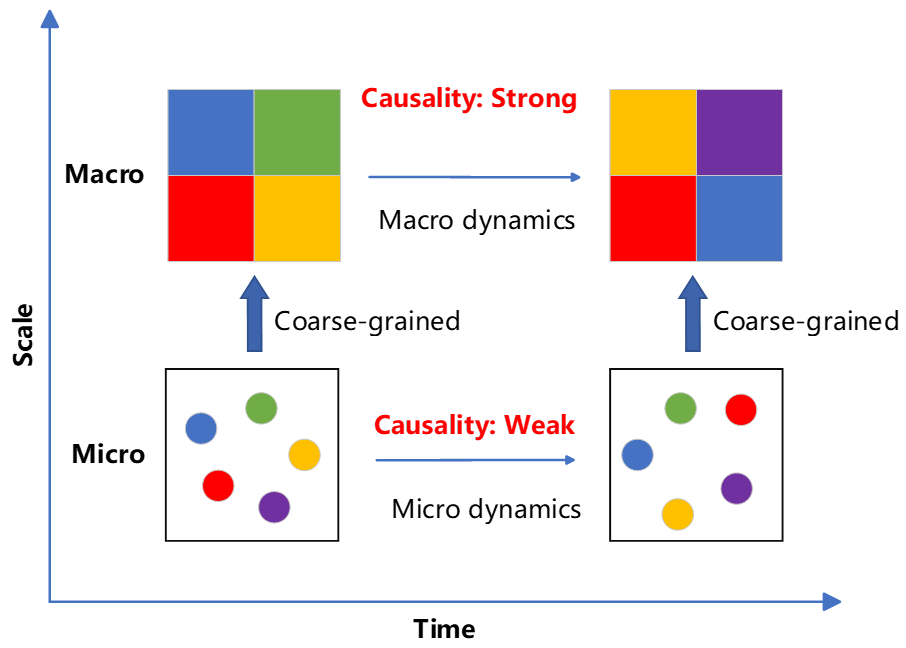}
\caption{Abstract Framework of Causal Emergence.}
\end{figure}

As shown in \textbf{Figure 3}, horizontal causation is primarily applied at the macro-level and aligns with the “dependency view” of causation. It examines causality from a correlational standpoint, aiming to determine whether a “cause-effect” relationship exists between two (or two types of) events, phenomena, or variables, and to evaluate the causal effect of the cause on the outcome. Vertical causation, on the other hand, belongs to the “generative view” of causation \cite{psillos2007causation}, which focuses on analyzing how the cause leads to the effect, through what pathways, under what conditions, and via what processes. It emphasizes uncovering the complex pathways and processes that link causes to outcomes. In the context of complex systems, it is often necessary to reconstruct the process behind emergent phenomena in order to truly grasp the underlying causal mechanisms.

Different perspectives on causality often correspond to different ways of "opening the black box". Horizontal causality holds that opening the black box means revealing the intermediate variables between treatments and outcomes. From this perspective, people should rely on quantitative tools to prioritize the identification of non-spurious relationships, establish counterfactual claims, and estimate unbiased quantitative parameters of these relationships when possible—all to infer from samples or test populations to unobserved target populations. Structural Equation Models (SEMs) and causal graphs are typical representatives of horizontal causality \cite{pearl2009causality}. From this viewpoint, the quality of data and how to describe such data creatively are of great importance.

In contrast, vertical causality argues that opening the black box involves decomposing the system into its individual components and generating the macroscopic phenomena under study by reconstructing the dynamic interactions between these components. Based on this perspective, priority should be given to methods that provide credible narratives to explain observed patterns. This view maintains that dependencies do not constitute causal mechanisms themselves, but rather are manifestations of such mechanisms. In response to this viewpoint, simulation methods become powerful tools for investigating the complex dynamic details between events. In particular, ABM is regarded as a key tool for establishing causal mechanisms, as it can demonstrate that the dependencies under study can be derived from hypothetical dynamic narratives \cite{anzola2020causation}.

Lucas’s critique of causal reasoning in macroeconomics—specifically regarding the claim that inflation causes employment—provides a classic example for understanding the differences between these two perspectives \cite{lucas1976econometric}. Within the "horizontal causation" camp, data-driven approaches have become the trend, establishing the central role of interventional methods. This has led to the sophisticated application of statistical techniques, the proliferation of time-series econometric models, and the development of vector autoregression (VAR) methods \cite{sims1980macroeconomics}.

In contrast, the "vertical causation" camp has embraced theory-driven approaches as its mainstream, requiring macroeconomic models to be integrated with "microfoundations." Initially, this led to the widespread adoption of rational choice models, which calculate economic aggregates based on individual preferences and expectations \cite{kirman1992whom, hoover2008does}. Subsequently, ABM was introduced to address issues related to agent heterogeneity and the aggregation of social networks \cite{tesfatsion2002agent, tesfatsion2006handbook, arthur2021foundations}. Sociology—analytical sociology in particular—has followed a similar path: shifting from regression-based statistical methods for survey data to ABM, in order to tackle micro-to-macro problems when interdependent structures exist \cite{manzo2020agent}.

Scholars who endorse the perspective of horizontal causation tend to be skeptical of the uniqueness of agents as a modeling tool for social mechanisms. For instance, Morgan argues, "The originality based on mechanisms has been overstated.... The claim that mechanisms are connections of nonlinear functions does not serve as a valid argument for rejecting the use of variables, since the basic elements of input-output functions can be redefined as variables" \cite{morgan2005edge}. This statement deserves special attention: it can be used to argue that, given that the theoretical formulation of mechanisms requires variables and functions, a set of structured intervention/mediator variables (i.e., mechanisms in the horizontal sense) can be regarded as "mechanism sketches," and multivariate statistics can serve as a direct tool for testing mechanism-based explanations.

From the perspective of vertical causation, however, this implication is incorrect, because the role of variables and functions in statistical models differs from their role in ABM. In the former case, variables and functions are used to detect patterns of average effects—patterns that may reflect the overall statistical characteristics of the hypothesized underlying mechanisms (composed of entities, attributes, and interactions). In contrast, in the latter case, variables and functions are employed to precisely represent the attributes, activities, and interactions of relevant micro-level entities. Furthermore, functions are used to initiate (i.e., simulate) the dynamics associated with the hypothesized mechanisms, with the goal of assessing under what conditions such dynamics give rise to the relevant macro-level phenomena.

Both horizontal causation and vertical causation possess irreplaceable advantages and inevitable shortcomings. Horizontal causation enables counterfactual verification through data analysis, yet it fails to illustrate the process by which causes lead to effects. In contrast, vertical causation can demonstrate the process of causes resulting in effects, but it struggles to obtain verification from real-world data. Horizontal causation can provide directional guidance for the experimental design of vertical causation. Such experimental design not only allows generativeness to be reflected in correlational causal relationships but also ensures the stability and verifiability of causal mechanisms. Meanwhile, vertical causation can offer mechanistic explanations for horizontal causation. Specifically, conclusions drawn from horizontal causal analysis can be deduced through experiments, capturing temporal dependencies and nonlinearities within causal structures—thereby providing a potential explanation for emergent phenomena at the system level.

\subsection{Forward Causation and Reverse Causation}
The concepts of “forward causation” and “reverse causation” proposed by Gelman \cite{gelman2011causality} distinguish between two fundamental types of causal inquiry. When one pursues a forward causal question, the aim is to understand and quantify “what will happen if we do X”; in contrast, interest in reverse causation involves asking “what caused Y.” In the former case, researchers focus on a particular phenomenon (e.g., education) and seek to identify the consequences of its presence, absence, or variation (e.g., its impact on fertility). In the latter case, attention is drawn to an observed outcome, and the goal is to retrospectively trace the various phenomena that may have led to this outcome.

The potential outcomes framework is widely used by statisticians and economists to address questions of forward causation or causal effects, with experiments being regarded as the canonical method for tackling this type of inference \cite{dawid2014fitting}. Once the effects of a given cause are established, a further question arises: why did the intervention produce the observed outcome? This leads to what is known as causal explanation \cite{sampson2013translating}. To answer this, one must return to reverse causation, designing a set of possible causes that could have produced a specific effect. This form of causal explanation attempts to open the black box of first-order causal dependencies through an interventionist approach. When the objective is to transform intervention effects into practical measures, it becomes essential to understand why an intervention works, whether the intervention’s effectiveness is context-dependent, and how it may vary over time. Gelman and Imbens argue that asking “why” questions can provide explanatory hypotheses for understanding causal effects and help identify flaws in assumed causal models \cite{gelman2013ask}.

Goldthorpe further clarified that causal relationships can be understood in at least three distinct ways: robust dependence, manipulability of outcomes, and generative processes \cite{goldthorpe2001causation}. In the first view, a causal claim depends on whether X continues to affect Y after introducing a set of other variables (denoted as Z) into the analysis. This perspective sees causality as contingent on controlling for confounding variables. Methods aligned with this view include time-series analysis, early causal path analysis, SEM, and more broadly, regression-based multivariate analytical techniques used in survey data analysis. In the second view, causality is interpreted as follows: if a causal factor X is manipulated, then under appropriate controls of confounding variables, a systematic effect on the response variable Y will be observed. This manipulative view of causality is aligned with the methodology of randomized experiments. Lastly, when causation is understood as a generative process, the focus shifts to explaining the nature and validity of the relational processes on which causality depends. In this view, simulation methods are considered promising tools for explaining underlying mechanisms.

The first scenario corresponds to a forward causal research paradigm, where researchers use structured or unstructured observational data—obtained from surveys, censuses, or online crawling data—to estimate counterfactual associations between independent variable X and dependent variable Y. Causality is defined as a robust, counterfactual association between observed variables. Researchers typically use statistical models to test and estimate these relationships. Common tools include propensity score methods (matching, stratification, weighting), instrumental variable techniques, path analysis, structural equation models, and directed acyclic graphs (DAGs) linking causes, mediators, and outcomes. In recent years, machine learning methods have also gained traction, including causal trees, causal forests, Bayesian additive regression trees, genetic matching, double robust methods, and generalized boosted models.

The second scenario also falls under the forward causal paradigm but integrates both longitudinal and cross-sectional perspectives. In longitudinal designs, a treatment group undergoes an intervention while a control group does not, allowing researchers to observe how Y changes in response to a change in X, relative to when X is unchanged. This thus helps infer causality between X and Y. Since complex systems often involve phenomena at the system-level rather than individual-level, such interventions must be repeated across multiple units to test generalizability. From the cross-sectional perspective, researchers collect pre- and post-treatment data for both groups and apply statistical models to test causal hypotheses and quantify effects—for instance, using difference-in-differences (DiD) or regression discontinuity designs to evaluate policy effects. However, due to practical constraints, field experiments in social science contexts are relatively rare. With the rise of the internet, online experiments have emerged as a viable alternative.

The third scenario belongs to the reverse causal research paradigm, characterized by a longitudinal, generative perspective, and often operationalized through ABM. In this framework, Y denotes the social phenomenon to be explained, X refers to a set of agents with diverse heterogeneous attributes, and rules specify agent behaviors and interaction mechanisms. Simulations illustrate how micro-level interactions give rise to specific macro-level phenomena. The actions and interactions of agents are the causes of emergent macro patterns. Tracing how agent-level behaviors and relationships “step-by-step” lead to these patterns provides a mechanistic explanation of the observed outcomes.

However, since ABM parameters are entirely defined by the researcher, questions often arise about external validity, i.e. the extent to which the model reflects real-world mechanisms. Moreover, simulation research faces a tension between model simplicity and realism: that is, overly simple models risk being dismissed as thought experiments, while overly complex models may better mirror reality but impede abstraction and causal generalization.

\section{Methodological framework of computational experiments}
\subsection{Computational Experiment = ABM + Experiment}
In response to the challenges faced by ABM in the field of causal inference, Professor Fei-Yue Wang systematically proposed the basic ideas, concepts and methods of computational experiment method in 2004, emphasizing the circular feedback relationship between artificial systems and actual systems \cite{Wang2004ParallelSM, Wang2004Computational, zhang2018cyber}. In recent years, Professor Xue and his team have carried out extensive and systematic work on computational experiment methods \cite{xue2018social, zhou2022sle2, zhou2024hierarchical, xue2023chatgpt, zhou2025federated, xuan2023self, xue2019analysis, xue2016computational, xue2016computationalER, xue2019analysis, xue2016Acomputational}. They provide a general framework for computational experiment method, which can integrate ABM logic and experimental techniques, and support the customization of general method for different application fields \cite{xiao2023putational, xue2023computational, ma2024computational}. The framework of computational experiment method includes five steps: artificial society modeling, experimental system construction, computational experiment design, computational experiment analysis, and computational experiment verification, and forms a feedback closed loop \cite{xue2024computationalA, xue2024computationalB, xue2023computationalB, lu2021computational, xue2021computational}.

Computational experiment methods are an improved and extended version of ABM. ABM is not simply used as a simulation tool, but as an “artificial social laboratory” that “spontaneously grows and cultivates” an alternative version of the actual system, and can conduct various “counterfactual experiments” on system behavior and decision analysis. On the one hand, computational experiments use ABM to emphasize vertical causality \cite{woodward2005making}, pay attention to the causes and formation process of phenomena, and give causal significance to the cross-level correlation information generated by ABM (the connection between micro-mechanisms and macro-emergence). On the other hand, computational experiments “use counterfactual algorithms to construct multiple worlds” to emphasize horizontal causality, establish a probabilistic relationship between model parameters and macro-emergence through experimental mechanisms, reveal the patterns and laws within the system, and provide “reference”, “prediction” and “guidance” for the possible operation of the actual system. Because the virtual foundation on which the experimental mechanism is based, that is, the attributes, behaviors and local environment of the agent, is a replica of its real-world counterpart. Therefore, the probabilistic relationships established between different levels of analysis can be viewed as counterfactual connections across levels of analysis in the real world.

\begin{figure}[htbp]
\centering
\includegraphics[width=\linewidth]{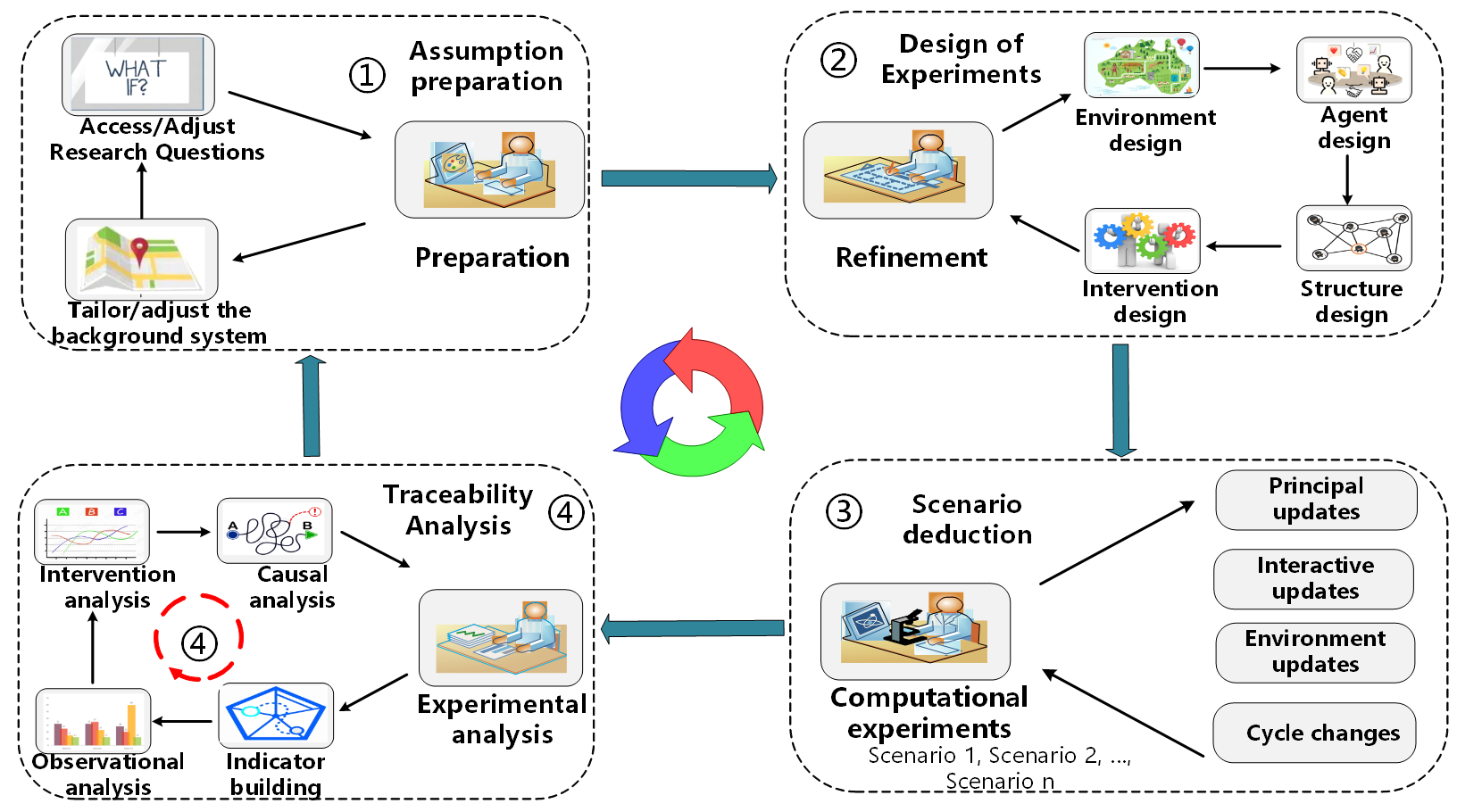}
\caption{The process model of the computational experiment method.}
\end{figure}

Computational experiments belong to the research path of “generative explanation”, that is, to trace the causes by exploring the mechanisms that can show how social phenomena arise \cite{Yuan2020Generative, epstein2012generative}. Traditional inductive interpretation uses statistical methods to derive general rules from data, and these rules are used to process data and are independent of data. However, generative explanations are based on the use of micro-interaction rules to derive macroscopic emergent structures, that is, general laws necessarily depend on the basis on which they are generated. Thus, generative explanations methodologically offer the idea that “how we know” is the basis of “whether we know”, a connection that is precisely what inductive and deductive methods lack. Generative explanations are different from traditional deductive or inductive paths, and place individual differences, local interactions, and the evolution of rules, which are ignored by traditional equilibrium analysis, at the core of the discussion. As shown in \textbf{Figure 4}, the computational experiment method is an “experiment + deduction” mechanism to explore the dynamic patterns, which consists of the following four core steps.
\subsubsection{\textbf{Preparation}}
The first step in a computational experiment is preparation, and the artificial society model can only be used for the computational experiment if it is proven to be consistent with actual scenario. In the process of research, the artificial society model needs to be continuously adjusted, and occurred situations can be used as a benchmark for model adjustment, similar to the standard sample in chemical experiments, until the simulated situation and the real situation match in terms of statistical distribution characteristics. In order to explore as many unknown scenarios as possible in the real world, it is necessary to analyze the conditions under which these unknown scenarios occur and the likelihood of these conditions, including: (1) an appropriate number of agents, with neither all agents nor some key agents ignored; (2) clear boundaries, abstracting the random perturbation factors that affect the state and properties of the system into boundaries, and eliminating the rest of the irrelevant parts; (3) agent-agent, agent-environment, and environment-environment interactions are based on knowledge rather than data, i.e., those linear/nonlinear, deterministic/random, static/dynamic function descriptions are based on the abstract relationship of facts; (4) eigenvectors are set up to represent the state of the system based on the research objectives, and these variables will be combined according to a certain relationship to achieve a measurement of the state of the system.
\subsubsection{\textbf{Design of Experiments}}
Computational experiments take the “multi-world” view of complex systems research, where the results of the experiment are treated as a possible “reality” that behaves “differently” but “equivalently” from reality. Experimental design follows an experimental methodology, including the setting of initial rules, the testing and selection of models, either to reproduce some real-world settings or phenomena, or to construct and observe possible worlds. Generative experiments deal with multivariate realizable cases: what if multiple microscopic mechanisms generate macroscopic patterns P (or rules) to be explained? The role of “experiments” is to assess the sufficiency of the generation of microscopic mechanisms: if multiple microscopic mechanisms can derive P, they can only be used as candidate explanations, and the most reliable explanation needs to be determined at the microscopic level. Therefore, it is necessary to continuously deduce according to the experimental design scheme, and the whole tracing process is a process of continuous experimentation. Further, the same initial setup may evolve into different possible outcomes. By designing counterfactual scenarios that have never happened in reality, it is possible to identify potential adverse effects that may occur and provide “references”, “predictions”, and “guidance” for possible scenarios.
\subsubsection{\textbf{Scenario Deduction}}
Computational experiment scheme deduction is deductive, which refers to the computational process of deriving macro rules from the initial configuration of the agents. ABM is particularly suitable for generative inference, which is to convert the generation process into an agent-based computational model, and reveal the micro-level generation mechanism of macroscopic rules through computational simulation. In this process, the ability to generate the macroscopic results of interest is key. Because each agent model is a computational program, its running process can be expressed by a loop function. Assuming the $n-th$ state of the system, then the $(n+1)$ state is deduced in a strictly deductive and computational manner. From the initial setup of the agents to every subsequent step, the computer implementation can be equated with the deduction of theorems. If $x$ is considered as the initial state of a set of agents, then it can be updated repeatedly according to the local interaction rules: if $x$, then $y$.
\subsubsection{\textbf{Mechanism-Based Explanation of Emergence}}
From the perspective of mechanism, the correlation and continuous co-variation between social phenomena and variables cannot explain social phenomena, but need to be explained by a series of agents and their behaviors that cause phenomenal changes. Mechanisms can help us understand why a particular cause is explanatory relevance. Something like: “Surprising fact C has been observed; if A is true, C is certainly true as well; therefore, there is reason to guess that A is true.” Here, hypothesis A is an attempted guess, and whether or not it can be recognized depends on A being able to explain Phenomenon C. So, the slogan of generative social science is: if you can't generate it, you can’t explain its emergence. In other words, the emergence of macroscopic outcomes depends on the micro-mechanism of local interactions, and this sufficiency of generation is a necessary condition for explaining how macroscopic outcomes are generated.

\subsection{Design Framework of Computational Experiments Method}
Compared with traditional experimental approaches, the computational experiment method offers a unique way to integrate descriptive, explanatory, and predictive models through generative deduction and generative experiments \cite{xue2024computationalA}. Its core idea is to explore the generative mechanisms of social phenomena rather than to explain them through general social laws or to merely identify statistically relevant factors. As shown in \textbf{Figure 5}, the comprehensive framework of computational experiment design consists of three modules: (1) \textbf{Descriptive Module} (upper part of Figure 7.3), which identifies influencing factors and response variables of dynamic system operations. (2) \textbf{Explanatory Module} (both sides of \textbf{Figure 5}), which observes the behavior of artificial societies under interventions (such as new strategies or parameter changes) through factorial experiment design \cite{salganik2006experimental}. (3) \textbf{Predictive Module} (lower part of \textbf{Figure 5}), which constructs a meta-model to provide deeper insights into the evolution of artificial societies. Details are as follows:

\begin{figure*}[h]
\centering
\includegraphics[width=16cm]{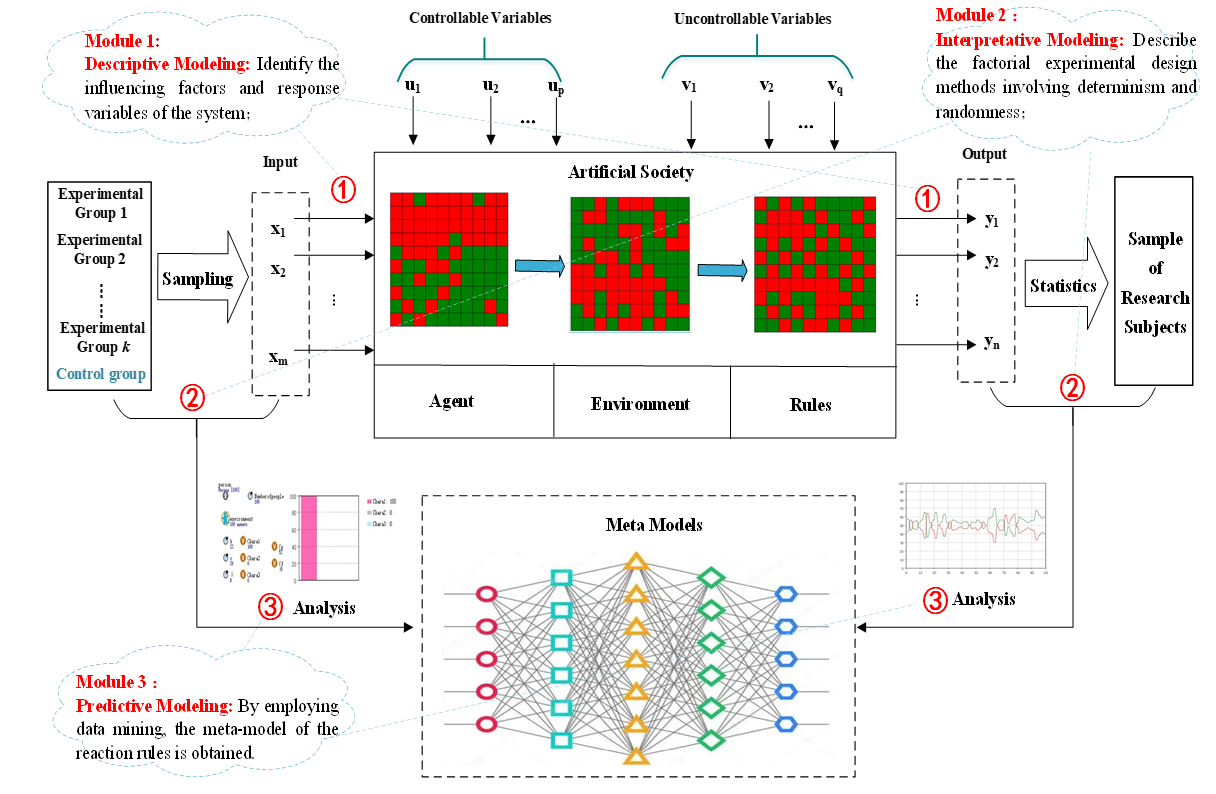}
\caption{The comprehensive modeling framework based on computational experiment design.}
\end{figure*}

\subsubsection{\textbf{Descriptive Module: Identification of Influencing Factors and Response Variables}}
Computational experiments employ computer simulation technology to address the generative explanation of complex social systems. The key question is: how do distributed local interactions among heterogeneous agents generate predetermined macro-level rules from the bottom up? Generative problems focus on “how” rather than “why,” and they are addressed by transforming generative processes into agent-based artificial societies. The key in this process is whether the artificial society can generate meaningful macro-level outcomes, otherwise the explanation fails. Generative sufficiency is a prerequisite for explaining the emergence of macro-level outcomes.

Artificial societies adhere to the principle of “simple consistency,” offering holistic descriptions of real-world complex systems \cite{xue2018social, peng2023computational}. Essentially, an artificial society is a formal logical model that investigates the relationships among a limited set of variables while controlling for others. As more variables are considered, accurately depicting the dynamics of an artificial society becomes increasingly challenging. On the one hand, events often result from the combined effects of multiple factors; on the other hand, the complexity and uncertainty of these factors complicate the characterization of their interrelationships. To facilitate subsequent explanatory and predictive modeling, the following steps should be followed: (1) Identify influencing factors: Which factors affect the system's state and behavior? (2) Analyze the paths and modes of influence: How do these factors exert their effects on the system? (3) Explore underlying factors: What influences these factors?

In general, unknown contexts can be categorized into several scenarios: (1) Abrupt environmental changes, especially those caused by major disturbances; (2) Changes in the network topology of agent interactions, such as fully connected graphs, random graphs, scale-free complex networks, small-world complex networks, among others; (3) System changes under intervention strategies, which are particularly common in social systems—for example, the introduction of new technologies, institutions, or cultures often requires ex-ante evaluation; (4) Changes in agent types, attributes, or behaviors—if the system is open, new agents may emerge, leading to contextual shifts; (5) Contextual variations due to changes in system noise; (6) Miscellaneous cases. As mentioned, regardless of the scenario, we must consider not only the contexts themselves but also their underlying causes and associated probabilities before conducting experiments. Once contexts are defined, researchers should not focus excessively on whether experimental conclusions align with prior experience. Many researchers make a mistake at this stage: when encountering results inconsistent with prior experience, they prematurely conclude that the experiment is invalid and abandon further analysis of these anomalous phenomena. Thus, they adhere rigidly to “heredity” while neglecting “variation.” Yet heredity bears little relevance to innovation, whereas variation constitutes its very essence.

\subsubsection{\textbf{Explanatory Module: Factorial Experiment Design}}
As shown in \textbf{Figure 5}, the experimental process can be viewed as a transformation of multiple input combinations into outputs with one or more observable response variables. Here, $x_1, x_2, …, x_m$ denote the inputs of the artificial social system; $y_1, y_2, …, y_n$ represent the outputs; $u_1, u_2, …, u_p$ indicate controllable factors or decisions; and $v_1, v_2, …, v_q$ correspond to uncontrollable factors or events. The objectives of experimental design are as follows: (1) To determine the set of factors influencing system outputs through computational experiments; (2) To identify the most effective controllable factors $u_i$ that can steer outputs toward optimal levels; (3) To select controllable factors $u_i$ that minimize the impact of uncontrollable factors or events $v_i$ on the system.

The most basic method to uncover relationships between inputs and outputs is manual model exploration guided by analysts or experts. Although this process is straightforward and requires no advanced technical expertise, it often introduces biases and may overlook potentially insightful unexplored regions of the input space. Therefore, a more systematic approach is needed to uncover input-output relationships: (1) For physical systems with deterministic elements and finite value domains, classical experimental design methods (e.g., random sampling, full or partial factorial designs, central composite designs) and modern methods (e.g., group screening designs, space-filling designs) are typically employed \cite{n2017design}. (2) For complex social systems with numerous uncertainties and random value spaces, Monte Carlo and resampling methods play crucial roles.

To develop computational experiment schemes and effectively guide experiments, in-depth research on computational experiment design is required. In comparison with traditional experiments, computational experiments exhibit the following characteristics: (1) All experimental factors can be easily modified, enabling experiments with a large number of factors and levels; (2) They can explore richer and more valuable insights within a larger experimental domain; (3) Pseudo-random numbers can be employed, and experiments remain controllable, thereby eliminating the need for randomization and blocking; (4) They place greater emphasis on sequential design, in which subsequent experimental points are determined based on previous results so as to effectively reduce sample size, avoid over-sampling, and improve simulation efficiency. Relative to general simulation experiments, computational experiments on complex systems are also characterized by a greater number of factors, more factor levels, and longer experiment runtimes.

\subsubsection{\textbf{Predictive Module: Meta-Models of Artificial Society}}
After collecting input and output data, in-depth analysis can extract the underlying rules of these relationships. The primary goal of this analysis is to improve the capacity to predict and control these relationships effectively. To overcome the limitations of manual exploration and experiment design-based techniques, the machine learning field provides a broad set of methods. These methods construct nonlinear mappings from artificial society inputs to outputs. The learned (potentially nonlinear) function (i.e., meta-model) serves as an approximate representation of the input–output relationship of the original simulation model \cite{ferber1998meta}.

Meta-model methods aim to directly predict outcomes of interest without explicitly focusing on identifying causal relationships. Traditional predictive models assume that out-of-sample data follow the same statistical distribution. In contrast, the focus here is on improving the applicability of predictive models across various contexts, such as uncontrollable factors or intentional interventions. It also encompasses extreme scenarios in entirely new situations where input features are set to unprecedented values. By comparing meta-model predictions with patterns observed in artificial society evolution, the accuracy and reliability of predictions can be reliably validated. Additionally, integrating optimization techniques (e.g., exact or heuristic methods) with meta-modeling offers a promising approach for identifying input parameter combinations that maximize or minimize specific characteristics of the original model outputs.

\subsection{Analytical Framework of Computational Experiments}
Understanding causal mechanisms in complex systems requires not only identifying causal effects themselves but also explaining the underlying causal processes. Computational experiments serve as a bridge connecting these two types of causal analysis. By simulating the operation of an “artificial society,” computational experiments can not only predict causal effects between system variables (i.e., forward causation) but also explore potential system causal mechanisms (i.e., reverse causation) by tracing simulated trajectories and counterfactual paths. This dual capability makes computational experimentation uniquely valuable in causal research: on one hand, it provides algorithmic estimations of causal effects; on the other, it offers operational modeling tools for causal explanation.

This analytical power relies on the construction and manipulation of a World Model \cite{zholus2022factorized} --an abstract representation of the structural and evolutionary rules governing real-world systems. Such models fall into two categories: the first is an external, objective world model (e.g., the Sora model), which learns from data, video, or other information to simulate environmental state transitions with high precision; the second is an internal, agent-centered belief model, which encodes how agents understand and expect the world to behave. By embedding both types of models, computational experiments can simulate not only the dynamic evolution of the external environment but also agents’ decision-making processes under different cognitive assumptions.

Within this computational experimental framework, forward causal analysis is typically implemented through intervention simulations—by altering the value of a system variable and observing changes in system outputs, one can uncover robust causal networks that describe causal interdependencies within the system. In contrast, reverse causal analysis relies on counterfactual trajectory generation, constructing alternative pathways that are similar to the actual pathway but differ in key variable values, thereby allowing for the inference of latent causal mechanisms. These two forms of causal analysis complement each other in simulation processes, thus jointly revealing the operating logic of world models and the dynamics of behavioral evolution.

\begin{figure*}[htbp]
\centering
\includegraphics[width=16cm]{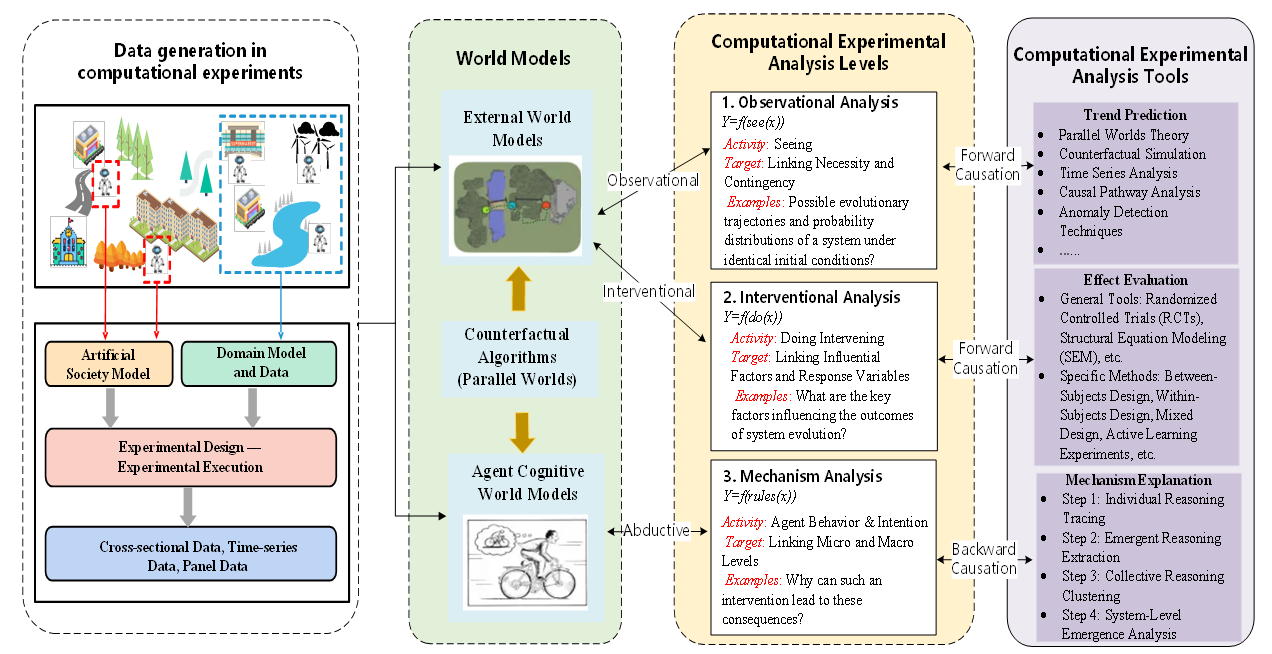}
\caption{Causal Inference Framework of Computational Experiments. The left side of the diagram illustrates the artificial society model operating in parallel with the real world, along with the ongoing data generation process. The middle section presents a three-layer analytical framework for computational experiments. The right side lists the tools and methods corresponding to each analytical layer.}
\end{figure*}

Based on this theoretical foundation, \textbf{Figure 6} presents a computational experimental analysis framework that explains how such simulations can be used to analyze world models within artificial societies. The framework consists of three layers of experimental analysis:

\begin{enumerate}
    \item The association layer (what is?): given a variable, what is the probability of an outcome occurring?
    \item The intervention layer (what if?): what would happen if we intervened or changed a condition?
    \item The mechanism layer (how is?): if a cause leads to an effect, what is the process through which it happens?
\end{enumerate}

The following sections will explore the specific techniques and methods used at each layer of this framework.

\subsubsection{\textbf{Observational Analysis (Linking Necessity and Contingency)}}
The first layer corresponds to a process based on an objective world model and aligns with the forward causation research paradigm. The core idea of observational analysis acknowledges the high uncertainty surrounding future outcomes and realization paths of events—manifested in the fact that the evolution of social systems involves multiple possibilities and is influenced by initial conditions and relevant external factors. Observational analysis focuses on modeling the possibility space of system evolution, identifying all potential states the social system might experience, along with their probability distributions. This process emphasizes the uncertainty in event evolution and uses scenario-based simulation data to reveal the randomness and unpredictability in system dynamics. Specifically, when facing multiple future possibilities, observational analysis seeks to quantify each trajectory and assess each trajectory’s likelihood within a probabilistic framework. This approach not only effectively highlights the role of contingency in system evolution but also provides theoretical support for preparing responses for various possible futures.

\subsubsection{\textbf{Interventional Analysis (Linking Cause and Effect)}}
The second layer also builds upon an objective world model and falls under the forward causation paradigm. Interventional analysis employs randomized controlled experiments to compare system trajectories across different hypothetical scenarios, observing how changes in intervention variables induce changes in outcome variables. This enables the identification of robust structural causal models. Key components of interventional analysis include:

\begin{itemize}
    \item Treatment/Control Groups: One scenario adopts an intervention A, while the control scenario does not; the differences in outcomes are then examined to determine whether intervention A caused changes in outcome B.
    \item Sample Consistency: Experimental and control scenarios are configured identically, ensuring that differences in outcomes are due to the intervention rather than other confounding factors.
    \item Random Sampling/Assignment: Initial parameters of both experimental and control scenarios are randomly sampled within specified ranges. By running parallel experiments for control and treatment groups, the Average Treatment Effect (ATE) can be used to quantify the intervention’s impact on the system.
\end{itemize}

\subsubsection{\textbf{Mechanism Analysis (Linking Micro and Macro)}}
The third layer represents the reverse causation research paradigm. This level involves not only analyzing the objective world model but also focusing on the agents’ subjective world models. Here, Y denotes the social phenomenon to be explained; X represents a set of agents with heterogeneous attributes; and Rule refers to the agent’s world model, revealing the internal cognitive structures and reasoning pathways that guide decision-making or actions. Mechanism analysis posits that agent actions and interactions are the root causes of macro-level phenomena, and that mechanistic explanations clarify how these micro-level dynamics lead to emergent patterns step by step. The internal world model of an agent functions like “turning the black box into a white box”: by analyzing behavioral trajectories and cognitive chains, one can understand how individual micro-level behaviors give rise to macro-level emergent outcomes. In this context, LLMs are powerful tools that can assist in uncovering the reasoning logic and behavioral intentions of agents.

\section{Case Study: Algorithmic Behavior on Internet Platforms}
With the development of mobile internet and the sharing economy, O2O (Online to Offline) service ecosystems have emerged as a new model for service element integration. These ecosystems enable interaction and integration among various components, including online platforms (such as Meituan, Didi, and HelloBike), offline entities, online payment systems, service providers (such as couriers, drivers, and delivery riders), and consumers. By 2023, the scale of employment on internet-based O2O platforms had reached an unprecedented level. Taking Meituan as an example, the number of registered delivery riders had surpassed 10 million. Internet platforms analyze data from workers (e.g., Meituan riders, Didi drivers) and feed the results back to these workers, making “Labor Order under Digital Control” possible. While this approach greatly improves labor efficiency, it has also given rise to a series of social issues, such as rider involution, algorithmic price discrimination (“data-enabled consumer exploitation”), and platform monopolies. This project adopts a “governance strategy + computational experiment” approach to study the O2O service ecosystem and uncover the underlying mechanisms of the platform economy. It also aims to explore how to design effective governance strategies that balance efficiency and fairness.

\subsection{Algorithmic Mechanisms of O2O Platforms}
The term “involution” was introduced by sociologist Alexander Gerschenkron to describe a state of stagnation or regression in socioeconomic development. Within social ecosystems, involution refers to excessive competition among individuals or groups in a resource-constrained environment, leading to decreased efficiency of input-output ratios and a decline in overall welfare. In this study, rider involution manifests as workers extending their working hours and increasing labor intensity in order to compete for a limited number of orders. However, their income does not increase significantly; instead, this behavior leads to deteriorating health conditions and growing income disparity. This research employs a LLM to construct a realistic multi-agent O2O platform simulation, aiming to analyze the causes and emergent dynamics of involution among delivery riders.

\begin{figure}[htbp]
\centering
\includegraphics[width=\linewidth]{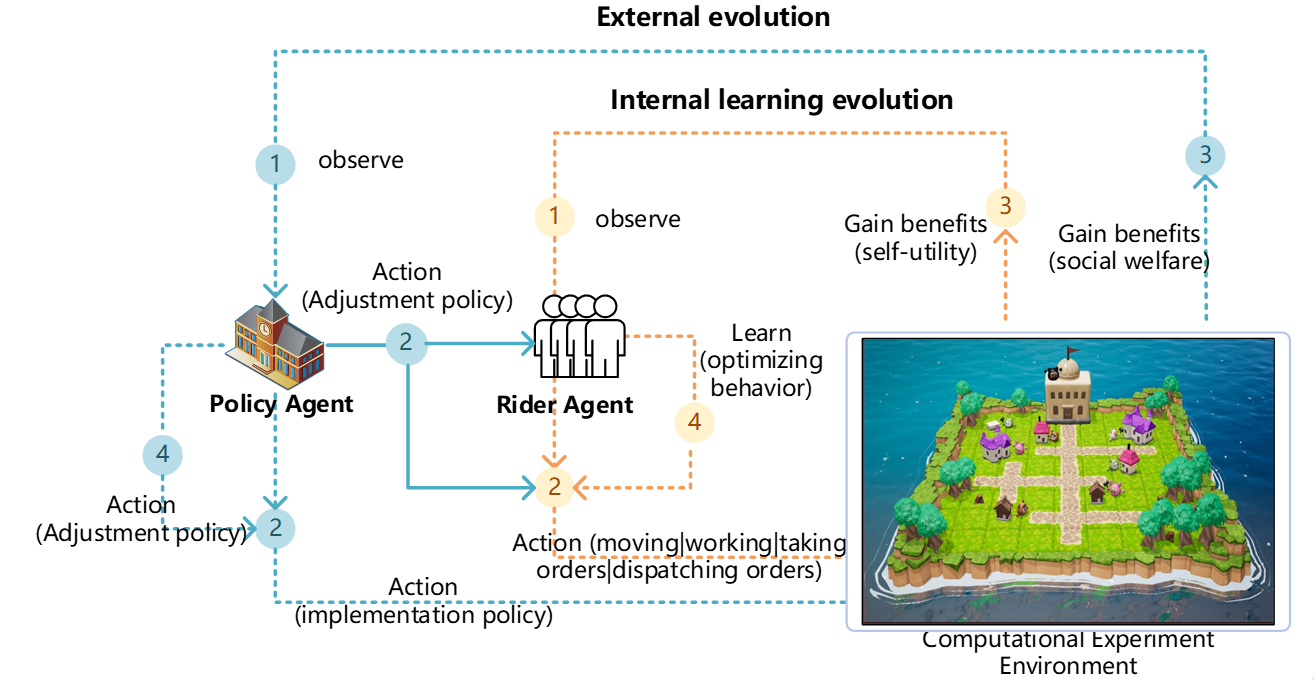}
\caption{Governance Strategy Optimization Framework Based on Two-Level Learning and Evolution.}
\end{figure}

We propose an experimental system based on a two-level learning-evolution framework, consisting of an inner and an outer loop. In this system, the Policy Agent represents the platform, dynamically adjusting governance strategies to balance individual welfare and overall system stability. The Rider Agent, representing riders, adapts its behavior through learning to maximize individual utility. A bi-directional feedback loop connects the platform algorithm and the agent system: riders’ learning behaviors affect system performance, which in turn drives the platform to adjust its policies. Through repeated iteration, the system evolves toward a dynamic equilibrium. \textbf{Table 2} presents the mapping between real-world delivery scenarios and the corresponding computational experiment system.

\begin{table*}[htbp]
\centering
\caption{Mapping Between the Real World and the Computational Experiment System.}
\renewcommand{\arraystretch}{1.3}
\begin{tabularx}{\textwidth}{|>{\raggedright\arraybackslash}p{1.5cm}
                               |>{\raggedright\arraybackslash}p{1.5cm}
                               |>{\raggedright\arraybackslash}X
                               |>{\raggedright\arraybackslash}X|}
\hline
\textbf{Model} & \textbf{Element} & \textbf{Real World} & \textbf{Computational System} \\
\hline

\multirow{2}{*}{\textbf{Agent Model}} 
& \textbf{Rider Agent} 
& Rider agents have indicators such as salary, speed, and unit consumption. They can independently choose or modify their operational area, within which they accept orders, and they also have the autonomy to determine their working hours. 
& The individual utility of a rider is defined as follows:

\[
Utility_{r,t} = \text{crra}\left( \sum_{t=0}^{t=T} Reward_{r,t} \right) - \sum_{t=0}^{t=T} Cost_{r,t}
\]

\[
\text{crra}(z) = \frac{z^{1-\eta} - 1}{1 - \eta}, \quad \eta > 0
\]

Where: Reward refers to the income earned by the rider; Cost represents expenditures, including time and distance. The CRRA model (Constant Relative Risk Aversion) is used to capture the cumulative payoff of an agent’s actions up to time $T$, reflecting the principle of diminishing marginal utility. \\

\cline{2-4}

& \textbf{Platform Agent} 
& During the rider’s labor process, the platform system is responsible for guiding their work, including order dispatching, rider matching, delivery pricing, and route planning. The platform also administers rewards and penalties to workers based on performance. The platform’s strategy optimization is driven by the objective of maximizing the social welfare function $Swf$, which requires balancing income equality and system productivity.
& The system utility function is defined as:

\[
Swf_t = eq_t(a_t^c) \cdot prod_t(a_t^c)
\]

Where: $a_t^c$ denotes the total income of riders in the system at time step $t$; $N$ is the number of riders in the system; $eq_t$ represents income equality; $prod_t$ denotes system productivity.

\[
eq_t(a_t^c) = 1 - \frac{gini(a_t^c) \cdot N}{N - 1}, \quad 0 \le eq_t(a_t^c) \le 1
\]

Where: 1 indicates perfect equality, and 0 indicates complete inequality.

\[
prod_t(a_t^c) = \sum_{i=1}^{N} a_{i,t}^c
\]

Where: $prod_t$ indicates the total income of all riders. \\

\hline

\multirow{2}{*}{\textbf{Rules Model}} 
& \textbf{Rider Agent Learning and Evolution} 
& If the price per delivery order decreases, riders must work longer hours, reduce the delivery time per order, and complete more deliveries in order to maintain their income. 
& The objective of Rider agent learning and optimization is to maximize individual utility per unit of time. Depending on the scenario, various machine learning approaches can be adopted, including LLM, evolutionary learning, and reinforcement learning (RL).
On one hand, agents can utilize imitation learning to dynamically adjust the weights of their utility functions based on environmental cues and the behavior of neighboring agents. On the other hand, agents can adopt RL- or LLM-based policies during decision-making and continuously improve their strategies through interaction with the environment. \\

\cline{2-4}

& \textbf{Platform Agent Learning and Evolution} 
& The platform’s strategy is continuously fine-tuned based on current system performance metrics to ensure its effectiveness. It updates delivery routes by analyzing rider trajectories and average delivery times, progressively shortening the guidance time allocated for each route. This process effectively constitutes a continuous test of user endurance and adaptability. 
& The degree of involution is measured by comparing system-level performance with individual-level utility.
$Involution_t = \frac{Swf_T}{\text{ave}(Utility_{T,i})}$

$Involution_t$: A larger value indicates more severe involution, meaning that as individual utility decreases while system-level performance remains stable, the degree of involution intensifies.\\
\hline

\textbf{Environment Model} 
& \textbf{Order Distribution}
& The city serves as the operational area for the riders. For the purposes of administrative management and statistical analysis, the entire city is divided into distinct zones. 
& A discrete grid map is used to represent the spatial environment of the riders, while orders are generated through Monte Carlo simulation. The generation formula is $f(x) = \sum_{i=1}^{5} a_i e^{-\left( \frac{x - b_i}{c_i} \right)^2}$
\\
\hline

\end{tabularx}
\end{table*}

\subsection{Experimental Objectives and Setup}
This study aims to investigate the phenomenon of involution among rider agents within an O2O experimental platform. The experimental objectives are threefold:

\begin{enumerate}
    \item RQ1: Can emergent phenomena be observed within the system?
    \item RQ2: Can we intervene to identify the key influencing factors of the system?
    \item RQ3: Can we uncover the underlying mechanisms behind the emergence to explain social dynamics through simulation?
\end{enumerate}

All experiments were conducted using a locally deployed DeepSeek-R1-Distill-Qwen-32B-FP8-Dynamic LLM, with the temperature parameter set to 0 to ensure repeatability. The experimental environment consists of a multi-agent system comprising five agent types: merchants, delivery riders, customers, government, and platform. For this study, we instantiated 100 rider agents.

The physical environment is a grid-based virtual map simulating a city with 10 commercial zones. Each rider agent’s reward function is designed to balance multiple objectives: minimizing working time, maximizing completed orders, and reducing labor costs. These objectives are weighted differently among riders to reflect varying decision-making priorities.

Rider agents are empowered by LLM-based intelligence, enabling autonomous learning and decision-making. Their decisions include selecting commercial zones, picking up orders, and executing random walk strategies. Throughout the 3600-step simulation—representing one month of activity—these rider agents continuously interact with the environment, leading to the dynamic evolution of the system.

To ensure the realism and validity of the experiments, the system was benchmarked using real-world data from the Zomato platform, which includes food delivery order records across multiple cities. The benchmark focused on simulating cross-city delivery scenarios, with key comparison metrics such as average effective working hours per rider per day and the relationship between working hours and order volume.

\begin{figure}[htbp]
\centering
\includegraphics[width=\linewidth]{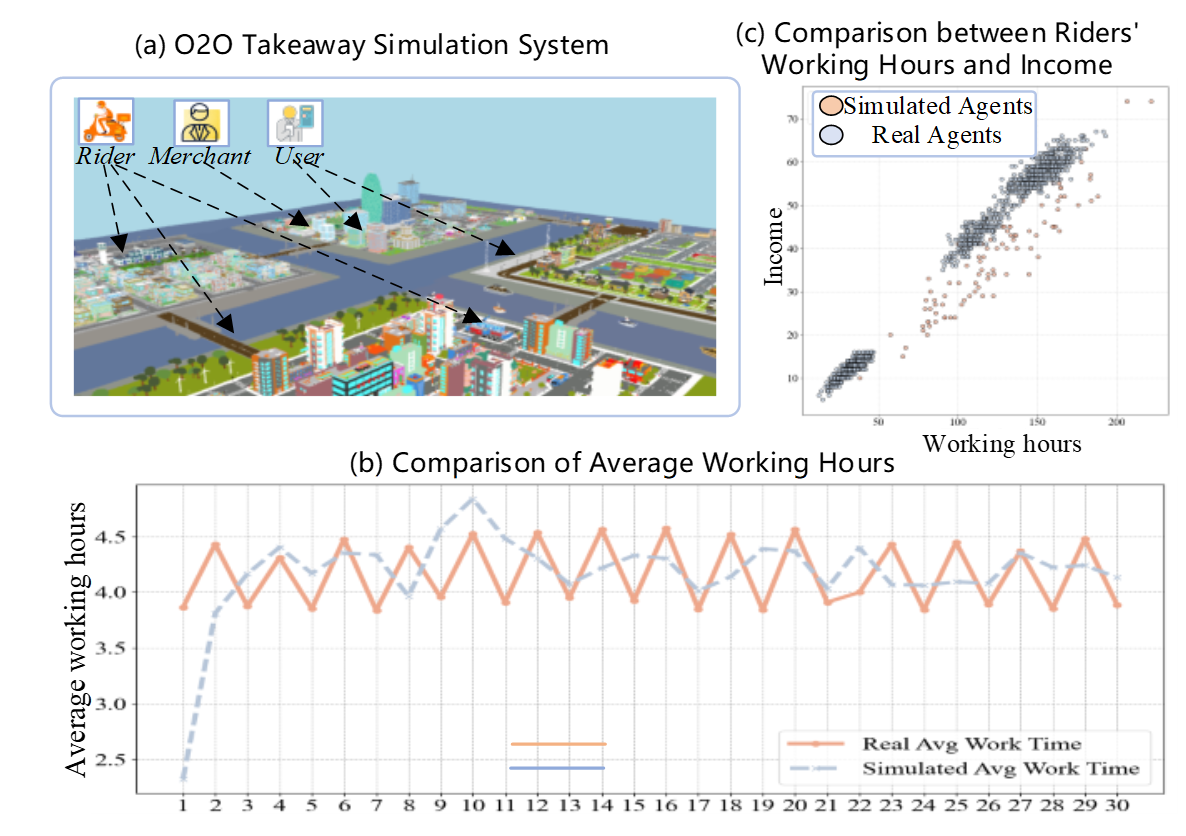}
\caption{Comparison between the Computational Experiment System and Real-World Datasets.}
\end{figure}

\textbf{Figure 8(b)} compares the average effective working hours per day, defined as the time spent actively delivering orders (excluding waiting and idle time), between the real-world dataset and the simulation results. The simulation yielded an average working time of 4.05 hours, which closely matches the real-world observation of 4.02 hours, with a Mean Absolute Error (MAE) of 0.05 hours and a Root Mean Square Error (RMSE) of 0.06 hours, indicating minimal deviation. Furthermore, the Pearson correlation coefficient of r = 0.98 demonstrates a strong linear correlation between simulated and real data.

These results collectively validate the accuracy and reliability of the proposed model in reproducing the distribution of real-world working hours.

\textbf{Figure 8(c)} illustrates the relationship between riders’ working hours and the total number of orders accepted. In general, longer working hours correlate with more orders completed. Gray dots represent data from 1,320 real riders, while orange dots denote data from the 100 simulated rider agents. The strong similarity between simulated and real data confirms the accuracy and robustness of the experimental setup.

\subsection{Experimental Analysis and Conclusions}
The experimental analysis consists of three levels: (1) Observational Analysis: This phase identifies potential involution phenomena based on an involution index and analyzes the behavioral trajectories of rider agents over a 30-day period. (2) Intervention Analysis: This phase tests the effects of three platform-level regulatory strategies: adjusting agent intelligence levels, optimizing interaction modes, and controlling order volume. (3) Mechanism Analysis: This phase investigates the underlying connections between agent intentions and behaviors, providing in-depth explanations of the emergent phenomena observed within the system.

\subsubsection{\textbf{Observational Analysis (RQ1)}}
This study adopts a repeated experimental approach to eliminate the influence of random factors. Under identical parameter settings, ten independent runs were conducted. Following industry benchmarks, the involution index (as defined in \textbf{Table 3}) was used to classify the level of involution into three categories: low involution (index $\leqslant$ 30); moderate involution (30 $<$ index $\leqslant$ 60); and high involution (index $>$ 60).

\begin{table}[htbp]
\centering
\caption{Comparison of Simulation and Actual Metrics}
\begin{tabular}{|l|c|}
\hline
\textbf{Metric} & \textbf{Value (h)} \\
\hline
Actual Average Working Time $T_{\text{real}}$ & 4.02 \\
Simulated Average Working Time $T_{\text{sim}}$ & 4.05 \\
Mean Absolute Error (MAE) & 0.05 \\
Root Mean Square Error (RMSE) & 0.06 \\
Pearson Correlation Coefficient $r$ & 0.98 \\
\hline
\end{tabular}
\end{table}

As illustrated in \textbf{Figure 9(a)}, the experimental distribution of involution levels shows that 75\% of the simulations exhibited a high level of involution under highly competitive conditions of the O2O platform. This indicates a strong consistency in the emergence of involution effects across repeated trials. \textbf{Figure 9(b)} further shows that the degree of involution tends to increase over time in all experiment types, eventually stabilizing at a relatively high level.

\begin{figure}[htbp]
\centering
\includegraphics[width=\linewidth]{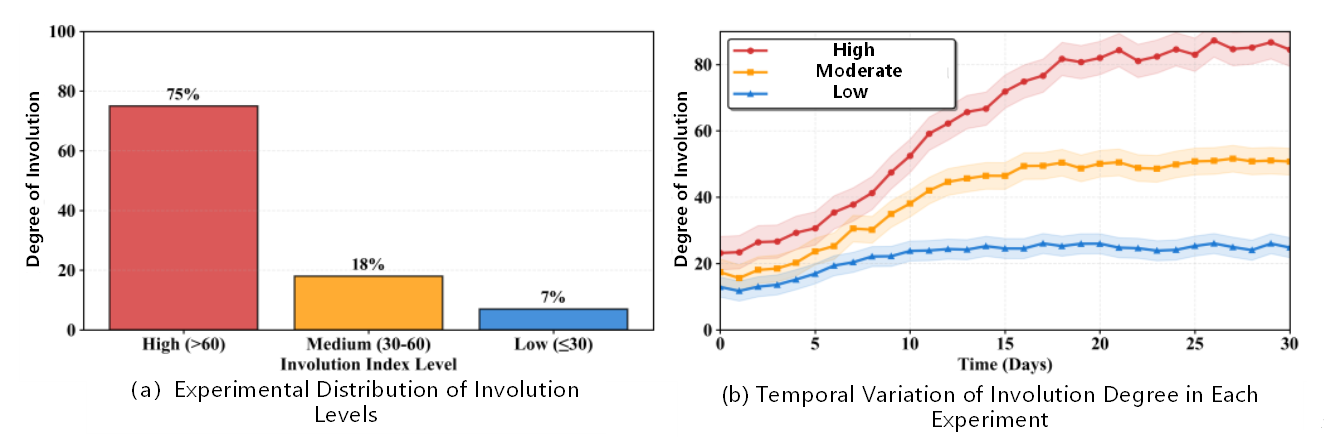}
\caption{Analysis of Involution Levels in the Experiment.}
\end{figure}

One outlier experiment exhibited a low level of involution, which significantly deviated from the overall trend. This anomaly can be attributed to three stochastic disturbances: a temporarily optimal initial spatial distribution of rider agents, pseudo-random variations in the path planning algorithm resulting in temporarily optimal routes, and a concentration of order assignments in low-density areas, which reduced competitive pressure.

As shown in \textbf{Figure 10}, we analyzed the temporal evolution of rider agents’ activity range in experiments characterized by a high degree of involution. In the heatmap, color intensity represents the density of rider distribution. Over time, the initially random spatial distribution of rider agents gradually evolves into concentrated clusters around locations such as merchants and users. This aggregation process visually reflects the emergence of involution behavior among rider agents.

\begin{figure}[htbp]
\centering
\includegraphics[width=\linewidth]{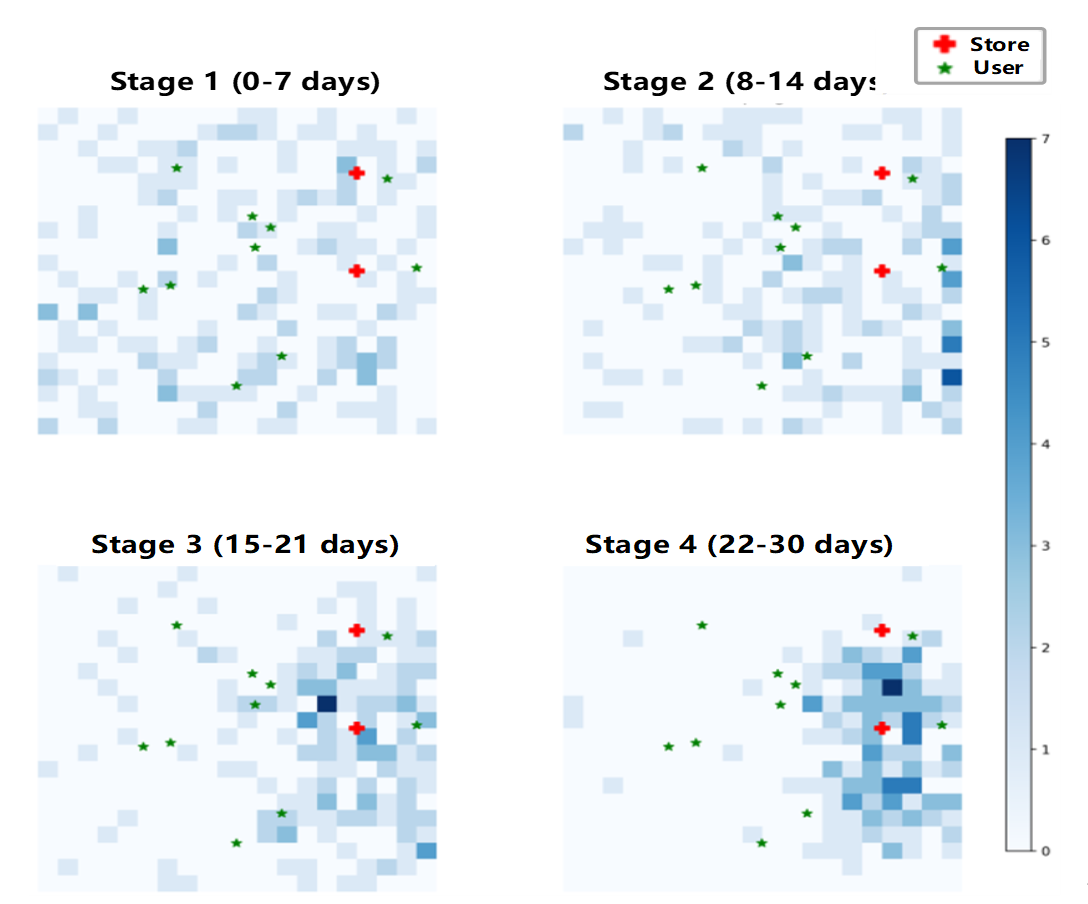}
\caption{Heatmap of Rider Activity Range.}
\end{figure}

\begin{figure*}[h]
\centering
\includegraphics[width=16cm]{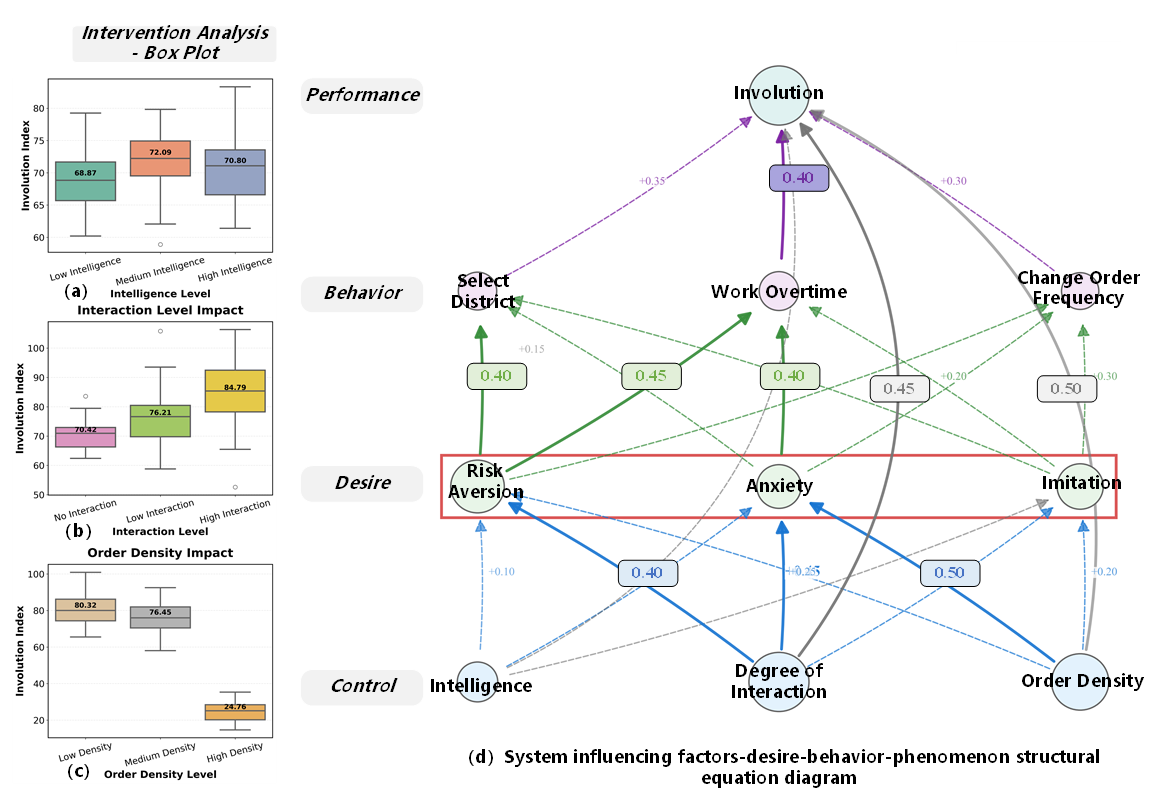}
\caption{Intervention Analysis Results.}
\end{figure*}

\subsubsection{\textbf{Intervention Analysis (RQ2)}}
To investigate the underlying mechanisms of involution, this phase of the study designed and conducted three targeted intervention experiments, each focusing on a specific influencing factor. The experiments explored the impact pathways of three core variables on involution dynamics: intelligence level adjustment (Factor A), interaction mode transformation (Factor B), and order volume fluctuation (Factor C). By applying multi-dimensional data analysis methods, the study aimed to clarify how these variables collectively influence individual and group-level performance within work scenarios.

As shown in the boxplot visualization in \textbf{Figure 11(a)}, the “intelligence level adjustment” intervention revealed no statistically significant differences in involution levels among the high-, medium-, and low-intelligence agent groups. All groups fell within the high involution range, with involution indices spanning from 68.87 to 72.09. This result indicates that under the current experimental setup, enhancing agent intelligence has limited effectiveness in mitigating systemic involution.

In contrast, \textbf{Figure 11(b)} highlights the pronounced heterogeneity in system behavior resulting from changes in interaction mode (Factor B). The introduction of a new interaction mechanism led to a significant increase in involution, with the involution index rising from 70.42 to 84.79 as interaction intensity increased. This suggests that higher levels of agent interaction may, in certain conditions, exacerbate involution rather than alleviate it.

Lastly, \textbf{Figure 11(c)} demonstrates that order volume (Factor C) exerts a strong influence on the degree of involution. As the number of available orders increased, the involution index decreased steadily from 80.32 to 24.76, indicating a shift from a high-involution to a low-involution state. This finding underscores the crucial role of resource abundance—in this case, sufficient order volume—in promoting stable system operations for riders and alleviating competitive pressure within the platform.

Further analysis of the relationship between influencing factors and the degree of involution, as shown in \textbf{Figure 11(d)}, was conducted using a SEM based on linear correlation calculations. The results reveal that interaction mode transformation and order volume are the two most critical factors affecting the experimental outcomes. Their respective standardized path coefficients are 0.45 and 0.50, indicating strong and direct impacts on the observed level of involution within the system.

\subsubsection{\textbf{Mechanism Analysis (RQ3)}}
Through clustering analysis of rider agents’ intention sequences (as shown in \textbf{Figures 12(a–c)}), the experiment identified several representative types of intentions. Over time, the clustering structure evolved from an initially homogeneous form into a more diverse, multi-cluster distribution. This temporal shift reflects the increasing differentiation of agent decision-making as the simulation progresses.

\begin{figure*}[!t]
\centering
\includegraphics[width=16cm]{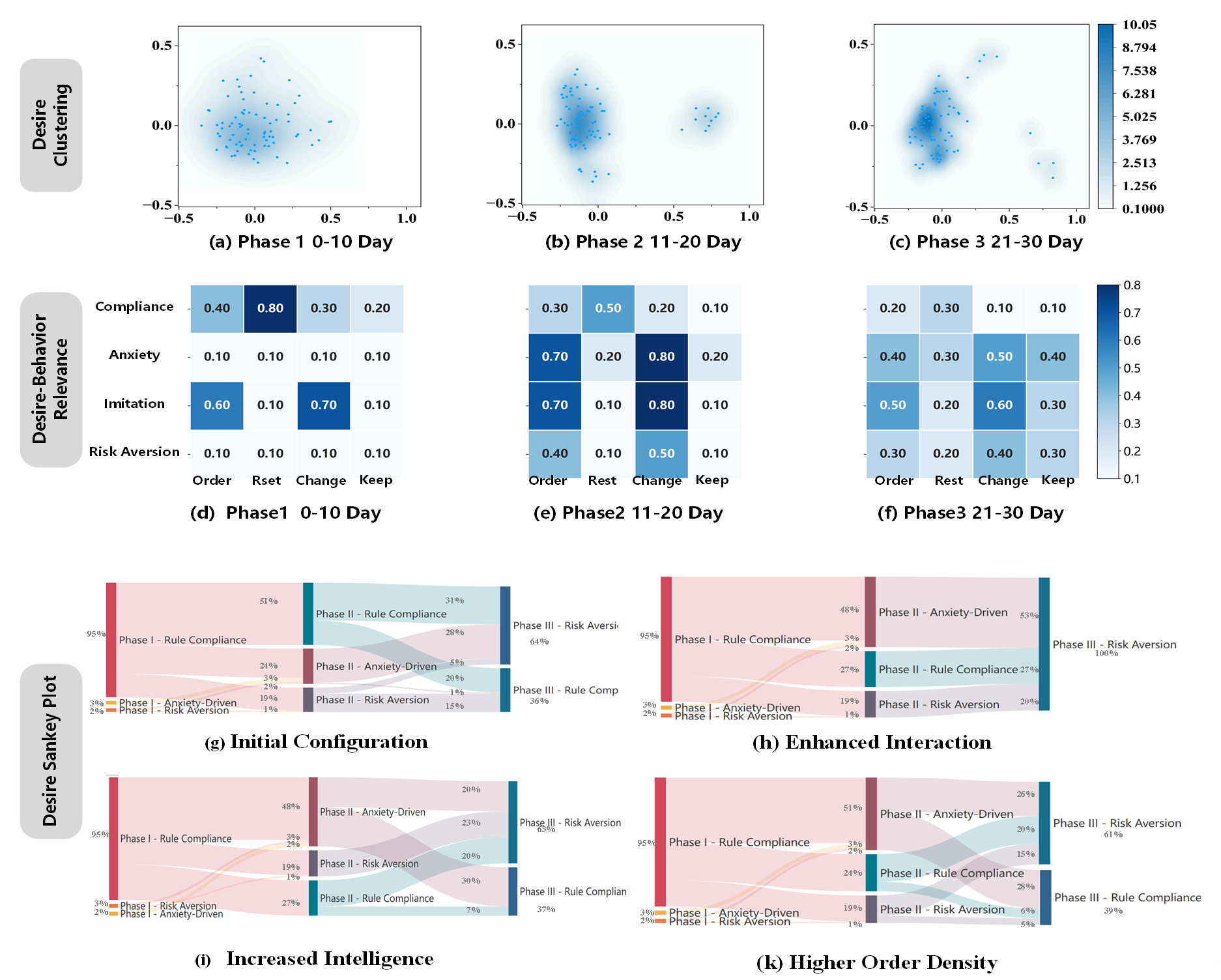}
\caption{Temporal Emergence of Intentions.}
\end{figure*}

Further correlation analysis conducted during the system’s temporal evolution reveals a notable relationship between agent intentions and behaviors in shaping involution dynamics. As shown in \textbf{Figures 12(d-f)}, factors such as business zone switching and order acceptance exhibit stronger correlations with intentions related to risk avoidance and anxiety, and these correlations grow more pronounced over time.

Notably, as highlighted in the red-marked region of \textbf{Figures 12(d)}, these two types of intentions are highly associated with typical involution behaviors such as business zone selection and extension of working hours. Specifically, the risk avoidance intention demonstrates a correlation coefficient of 0.40 with regional selection and 0.45 with work hour extension. These findings underscore the critical role of individual intentions in driving and sustaining involution behaviors within the system.

As shown in \textbf{Figures 12(g)}, a Sankey-based flow visualization method was used to trace the fine-grained transitions in agent behavior over a 30-day observation period. The decision-making intentions of rider agents underwent three distinct stages: an initial rule-following phase, followed around Day 15 by a rise in anxiety-driven behaviors (increasing to 24\%), and eventually transitioning to a dominant risk-avoidant phase, reaching 64\% by the end of the 30-day observation period. Correspondingly, behavior patterns evolved from rest-focused, to aggressive order grabbing, and finally to risk-avoidant strategies.

To uncover the mechanisms driving these patterns, three intervention scenarios were evaluated. In the “enhanced interaction” setting, agents overwhelmingly adopted risk-avoidant behaviors, indicating that deeper interaction intensity increases the tendency toward conservative decision-making, as shown in \textbf{Figures 12(h)}. In the “enhanced intelligence” scenario, the share of risk-averse intentions rose even further, suggesting that stronger cognitive capacity reinforces caution, as shown in \textbf{Figures 12(i)}. Conversely, in the “increased order density” scenario, the proportion of risk-aware agents dropped to 61\%, implying that higher order density-induced workloads reduce risk sensitivity and lead to more risk-taking behavior, as shown in \textbf{Figures 12(k)}.

These findings indicate that interaction dynamics and environmental pressure are the two key drivers of involution. Enhanced interaction depth significantly amplifies agents’ inclination toward risk avoidance, showing that changes in interaction modes directly affect decision-making styles. Meanwhile, higher order density—representing greater task loads—can suppress risk awareness and push agents toward competitive and aggressive strategies.

In summary, the emergence of involution is not driven by a single factor but is instead the result of the combined effects of interaction mechanisms, environmental stress, and cognitive ability. This conclusion aligns with the results of the observational and intervention analyses, highlighting the complexity of behavioral emergence in agent-based O2O systems. Summary of Findings Across the Three Analytical Layers:
(1)	Observational Analysis: Agents in the O2O system increasingly exhibit involution over time, with their behaviors converging and becoming more homogeneous.
(2)	Intervention Analysis: Timely introduction of governance strategies—such as order-balancing mechanisms—during the anxiety-dominated phase can effectively mitigate over-competition and enhance system performance and stability.
(3)	Mechanism Analysis: Order scarcity and anxiety diffusion trigger risk-avoidant intentions, which in turn drive cross-zone order grabbing and extended working hours, forming the core behavioral patterns that underlie rider involution.

\section{Summary}
Although computational experimentation has made significant progress, its development still faces numerous challenges. One of the key issues lies in bridging the gap between descriptive modeling (data analysis) and predictive modeling (meta-modeling). While scattered efforts have been made in this direction, there is still a lack of a systematic and mature framework for causal inference through computational experiments. To address this gap, this part proposes a multi-layer integrated framework that combines computational experiments with causal reasoning. The framework is designed to tackle the challenges of identifying and analyzing causal relationships in complex social systems, which is mainly composed of three levels: (1) Observational Analysis: Based on the data from computational experiments, explores causal relationships between influencing factors and response variables, and uses anomaly analysis methods to capture system anomalies. (2) Intervention Analysis: Implements various interventions to observe changes in the system, further evaluating causal effects. (3) Mechanism Analysis: By correlating micro-level behaviors with macro phenomena, explains how "interventions" induce "outcomes" through what pathways, under what regulatory conditions, and via what processes. Finally, the phenomenon of rider involution on food delivery platforms is used as a case study to effectively demonstrate the validity, generality, and efficiency of the proposed causal inference framework.

\section*{Acknowledgment}
This work has been supported in part by the National Natural Science Foundation of China (No.62472306, No.62441221,No.62206116), Tianjin University's  2024 Special Project on Disciplinary Development (No.XKJS-2024-5-9), Tianjin University Talent Innovation Reward Program for Literature \& Science Graduate Student (C1-2022-010), and Henan Province Key Research and Development Program (No.251111210500).

\bibliographystyle{IEEEtran}
\bibliography{2025.9.25}

\end{document}